\documentclass[twocolumn,twoside,slac_two]{revtex4}
\usepackage{graphicx}
\usepackage{fancyhdr}
\begin{document}

\title{$\Upsilon(1S)-->\gamma(\eta',\eta,f_2(1270))$ decays}

%

\author{B.A.Li}
\affiliation{Department of Physics and Astronomy, University of Kentucky, Lexington, KY 40506, USA}

\begin{abstract}
$\Upsilon(1s)\rightarrow\gamma(\eta',\eta, f_2(1270))$ are studied by an approach in which the glueball 
components of these mesons play dominant role.
Strong dependence on quark mass has been found in the
decays $(J/\psi, \Upsilon(1s))\rightarrow\gamma(\eta',\eta)$.
Very small decay rates of $\Upsilon(1s)\rightarrow\gamma(\eta',\eta)$
are predicted. d-wave dominance has been found in $\Upsilon(1s)\rightarrow\gamma f_2(1270)$, which leads to larger
$B(\Upsilon(1s)\rightarrow\gamma f_2(1270)).$
\end{abstract}

\maketitle

\thispagestyle{fancy}



Very small upper limits of the branching ratios of radiative $\Upsilon(1s)$ decays
into $\eta$ and $\eta'$ have been reported by CLEO[1]
\begin{eqnarray}
B(\Upsilon(1s)\rightarrow\gamma\eta)<1.0\times10^{-6},\nonumber \\
B(\Upsilon(1s)\rightarrow\gamma\eta')<1.9\times10^{-6}.
\end{eqnarray}
at the $90\%$C.L. $B(\Upsilon(1S)\rightarrow\gamma\eta')$ is smaller than
\begin{equation}
B(J/\psi\rightarrow\gamma\eta')=(4.71\pm0.27)\times 10^{-3}
\end{equation}
by almost three order of magnitudes.
On the other hand[2],
\begin{eqnarray}
B(\Upsilon(1S)\rightarrow\gamma f_2(1270))&=&(10.2\pm0.8\pm0.7)\times 10^{-5},\nonumber \\
B(\Upsilon(1S)\rightarrow\gamma f_2(1270))&=&(10.5\pm1.6^{+1.9}_{-1.8})\times 10^{-5}.
\end{eqnarray}
are reported by CLEO.
$B(\Upsilon(1S)\rightarrow\gamma f_2(1270))$ is about one order of magnitude smaller than
\begin{equation}
B(J/\psi\rightarrow\gamma f_2(1270))=(1.43\pm0.11)\times 10^{-3}.
\end{equation}
In this talk we try to understand why $B(\Upsilon(1s)\rightarrow\gamma\eta')$ is so small and
$B(\Upsilon(1S)\rightarrow\gamma f_2(1270))$ is not that small.

In pQCD radiative decay of a heavy vector meson is described as $V\rightarrow\gamma gg$.
The two gluons are color singlet and they can form $0^{++}, 0^{-+}$ and $2^{++}$ states.
$J/\psi$ radiative decay has long been regarded as a fertile hunting ground for glueballs.
So far, the existence of glueball has not been established. The efforts in search for glueball are complicated by
mixture between the components of gluons and quarks.
The difference between the radiative decays of $J/\psi$ and $\Upsilon(1s)$ is caused by the mass difference of c- and b-quark.
In 1984[3] we have studied $J/\psi\rightarrow\gamma\eta, \gamma\eta'$. In this study
$\eta'$ and $\eta$ are taken as mixing states of glueball and $q\bar{q}$ mesons and the gluon contents make
dominant contributions to $J/\psi\rightarrow\gamma\eta, \gamma\eta'$.
The ratio
\begin{equation}
\frac{\Gamma(J/\psi\rightarrow\gamma\eta')}{\Gamma(J/\psi\rightarrow\gamma\eta)}=5.30\pm0.05
\end{equation}
has been predicted, which agrees with data very well.
The same approach is applied to study $\Upsilon\rightarrow\gamma\eta'(\eta)$.
The U(1) anomaly of $\eta'$ meson shows that there is strong coupling between $\eta'$ and two gluons.
and $\eta'$ contains substantial gluon content.
In this approach the decays of $J/\psi\rightarrow\gamma\eta'(\eta)$
contain two parts: the amplitude of $J/\psi\rightarrow\gamma gg$ is calculated by pQCD and
the couplings $gg\rightarrow\eta'(\eta)$ are dominated by the gluon contents of $\eta'$ and $\eta$.
The flavor singlet part of $\eta$ and $\eta'$ is a mixing state of glueball and quark singlet
\begin{equation}
|\eta_0>=sin\phi|G>+cos\phi|q\bar{q}>,
\end{equation}
where $\phi$ is the mixing angle. It is argued in Ref.[3] that
the quark contents of $\eta$ and $\eta'$ are suppressed in $J/\psi\rightarrow\gamma\eta(\eta')$ by $O(\alpha_s^2)$.
Therefore, only the glueball content is taken into account in the calculations of $\Gamma(J/\psi\rightarrow\gamma\eta(\eta'))$.
The coupling between state-G and two gluons is expressed as
\begin{eqnarray}
<G|T\{A^a_\alpha(x_1)A^b_{\beta}(x_2)\}|0>&=&\frac{\delta_{ab}}{\sqrt{2E_G}}\nonumber \\
\epsilon_{\alpha\beta\mu\nu}(x_1-x_2)^\mu p^\nu f_G(0)e^{{i\over2}p_G (x_1+x_2)}.
\end{eqnarray}
The decay widths are derived as
\begin{eqnarray}
\Gamma(J/\psi\rightarrow\gamma\eta')=cos^2\theta sin^2\phi\frac{2^{11}}{81}\alpha
\alpha^2_s(m_c)\psi^2_J(0)f^2_G \nonumber \\
{1\over m^8_c}\frac{(1-{m^2_{\eta'}\over m^2_J})^3}
{\{1-2\frac{m^2_{\eta'}}{m^2_J}+{4m^2_c\over m^2_J}\}^2}\nonumber \\
\{2m^2_J-3m^2_{\eta'}(1+{2m_c\over m_J})-16{m^3_c\over m_J}\}^2,
\end{eqnarray}
\begin{eqnarray}
\Gamma(J/\psi\rightarrow\gamma\eta)=sin^2\theta sin^2\phi\frac{2^{11}}{81}\alpha
\alpha^2_s(m_c)\psi^2_J (0)f^2_G\nonumber \\
{1\over m^8_c}\frac{(1-{m^2_\eta\over m^2_J})^3}
{\{1-2\frac{m^2_\eta}{m^2_J}+{4m^2_c\over m^2_J}\}^2}\nonumber \\
\{2m^2_J-3m^2_\eta(1+{2m_c\over m_J})-16{m^3_c\over m_J}\}^2,
\end{eqnarray}
\begin{equation}
\psi^2_J(0)=\frac{27}{64\pi\alpha^2}m^2_J\Gamma_{J/\psi\rightarrow ee^+}.
\end{equation}
In the study of $J/\psi\rightarrow
\gamma+f(1273)$[4] it is found that $m_c=1.3GeV$ fits the data very well and this value is
in the range of
\(m_c=1.25\pm0.09GeV\)[5], and the mixing angle is taken
\(\theta=-10.75^0\pm0.05^0 .\)
The value(5) is predicted

The factor $\frac{1}{m^8_c}$ in Eq.(8) shows strong dependence of the decay rate on $m_c$.
Because of the cancellation between $m_J$ and $m_c$ in the factors 
\[\{1-2\frac{m^2_\eta}{m^2_J}+{4m^2_c\over m^2_J}\}^2,\]
\[\{2m^2_J-3m^2_\eta(1+{2m_c\over m_J})-16{m^3_c\over m_J}\}^2\]
of Eq.(8)
the decay rates are very sensitive to the value of $m_c$.

By changing corresponding quantities,
the decay rates of $\Upsilon(1s)\rightarrow\gamma\eta', \gamma\eta$ are
obtained. The ratio is determined to be
\begin{eqnarray}
R_{\eta'}=\frac{B(\Upsilon\rightarrow\gamma\eta')}{B(J/\psi\rightarrow\gamma\eta')}
={1\over4}\frac{\alpha^2_s(m_b)}{\alpha^2_s(m_c)}\frac{\psi^2_\Upsilon (0)}{\psi^2_J (0)}{m^8_c\over m^8_b}\nonumber \\
\frac{(1-\frac{m^2_{\eta'}}{m^2_\Upsilon})^3}{(1-\frac{m^2_{\eta'}}{m^2_J})^2}
\frac{(1-2\frac{m^2_{\eta'}}{m^2_J}+4\frac{m^2_c}{m^2_J})^2}
{(1-2\frac{m^2_{\eta'}}{ m^2_\Upsilon}+4\frac{m^2_b}{ m^2_\Upsilon})^2}\nonumber \\
\frac{\{2m^2_\Upsilon-3m^2_{\eta'}(1+{2m_b\over m_\Upsilon})-16{m^3_b\over m_\Upsilon}\}^2}
{\{2m^2_J-3m^2_{\eta'}(1+{2m_c\over m_J})-16{m^3_c\over m_J}\}^3}
\frac{\Gamma_{J/\psi}}{\Gamma_\Upsilon},
\end{eqnarray}
where
\begin{equation}
\frac{\psi^2_{\Upsilon}(0)}{\psi^2_J(0)}=4\frac{\Gamma_{\Upsilon\rightarrow ee^+}}
{\Gamma_{J/\psi\rightarrow ee^+}}\frac{m^2_{\Upsilon}}{m^2_J}.
\end{equation}
If \(m_J=2m_c\) and \(m_\Upsilon=2m_b\) are taken
\begin{eqnarray}
R_{\eta'}=\frac{B(\Upsilon\rightarrow\gamma\eta')}{B(J/\psi\rightarrow\gamma\eta')} \nonumber 
\end{eqnarray}
\begin{eqnarray}
R_{\eta'}=\frac{\Gamma(\Upsilon\rightarrow\gamma\eta')/\Gamma(\Upsilon\rightarrow light\; hadrons)}
{\Gamma(J/\psi\rightarrow\gamma\eta')/\Gamma(J/\psi\rightarrow\; light hadrons)}\times\nonumber 
\end{eqnarray}
\begin{eqnarray}
\frac{B(\Upsilon\rightarrow light\; hadrons)}{B(J/\psi\rightarrow light\; hadrons)}\nonumber
\end{eqnarray}
\begin{eqnarray}
=\frac{\alpha_s(m_c)}{\alpha_s(m_b)}({m_c\over m_b})^7\frac{1-{m^2_{\eta'}\over 4m^2_b}}
{1-{m^2_{\eta'}\over 4m^2_c}}\frac{B(\Upsilon\rightarrow light\; hadrons)}
{B(J/\psi\rightarrow light\; hadrons)}\times\nonumber \\
\frac{\Gamma_{\Upsilon\rightarrow ee}}{\Gamma_{J/\psi\rightarrow ee}}
=0.29\frac{\alpha_s(m_c)}{\alpha_s(m_b)}({m_c\over m_b})^7.
\end{eqnarray}
Comparing with all other studies, stronger dependence on quark masses and small coefficient
are obtained by this approach.
\(m_b=4.7\) GeV and \(m_c=1.3\)GeV are taken.
Inputting $B(J/\psi\rightarrow\gamma\eta')$, it is obtained
\begin{eqnarray}
B(\Upsilon\rightarrow\gamma\eta')&=&R_{\eta'} B(J/\psi\rightarrow\gamma\eta')
=1.04\times^{-7}\nonumber \\
B(\Upsilon\rightarrow\gamma\eta)&=&0.022B(\Upsilon\rightarrow\gamma\eta')=0.23\times10^{-8}.
\end{eqnarray}
Both branching ratios are less than the experimental upper limits.

The same approach has been applied to study $J/\psi, \Upsilon(1s)\rightarrow\gamma f_2(1270)$[4], in which the vector
bosons decays to $\gamma gg$ and $f_2(1270)$ is coupled to two gluons. The tensor meson $f_2(1270)$ contains
glueball components
\begin{equation}
|f_2>=cos \phi |q\bar{q}>+sin\phi |gg>.
\end{equation}
The glueball component of $f_2$ is
dominant in the decay $J/\psi, \Upsilon(1S)\rightarrow\gamma f_2$.
We have studied $J/\psi\rightarrow\gamma f_2(1270)$ long time ago[4].
\begin{eqnarray}
<f_{\lambda_2}\gamma_{\lambda_1}|S|J_{\lambda}>=(2\pi)^4\delta(p_J-p_{\gamma}-p_f)
e(8\omega_{\gamma}E_f E_J)^{-{1\over2}}
T_{\lambda_2},\nonumber\\
T_0=-{2\over\sqrt{6}}(A_2+p^2 A_1),\nonumber \\
T_1=-{\sqrt{2}\over m_J}(E A_2+m_f p^2 A_3),\nonumber \\
T_2=-2A_2,\nonumber \\
\end{eqnarray}
\begin{eqnarray}
A_1=-a\frac{2m^2_f-m_J(m_J-2m_c)}{m_cm_J[m^2_c+{1\over4}(m^2_J-2m^2_f)]},\nonumber \\
A_2=-a{1\over m_c}\{{m^2_f\over m_J}-m_J+2m_c\},\nonumber \\
A_3=-a\frac{m^2_f-{1\over2}(m_J-2m_c)^2}{m_c m_J[m^2_c+{1\over4}(m^2_J-2m^2_f)]},\nonumber \\
a={16\pi\over 3\sqrt{3}}\alpha_s(m_c) G(0)\psi_J(0){\sqrt{m_J}\over m^2_c},\nonumber \\
E={1\over2m_f}(m^2_J+m^2_f),\;\;p={1\over2m_f}(m^2_J-m^2_f),
\end{eqnarray}
where G(0) and $\psi_J(0)$ are the wave functions at origin of the glueball and $J/\psi$ respectively.
The decay width of $J/\psi\rightarrow\gamma f_2$ is derived as
\begin{eqnarray}
\Gamma(J/\psi\rightarrow\gamma f_2)=\frac{128\pi\alpha}{81}sin^2\phi\alpha_s^2(m_c)G^2(0)\psi^2_J(0)\nonumber \\
{1\over m^4_c}
(1-{m^2_f\over m^2_J})\{T^2_0+T^2_1+T^2_2\}.
\end{eqnarray}
The ratios of the helicity amplitudes are defined as
\begin{equation}
x={T_1\over T_0},\;\;y={T_2\over T_0}.
\end{equation}
It is obtained
\(y=0\) where \(m_c=1.3GeV\) is taken. 
It agrees with data.
By changing corresponding quantities, the helicity amplitudes and the decay width of
$\Upsilon(1s)\rightarrow\gamma f_2$ are obtained.
The ratios of the helicity amplitudes are obtained
\begin{equation}
x^2=0.058,\;\;y^2=5.9\times10^{-3}.
\end{equation}
They are consistent with experimental values[5]
\begin{equation}
x^2=0.00^{+0.02+0.01}_{-0.00-0.00},\;\;y^2=0.09^{+0.08+0.04}_{-0.07-0.03}.
\end{equation}
In order to show the quark mass dependence of the ratio, $\frac{B(\Upsilon\rightarrow\gamma f_2)}
{B(J/\psi\rightarrow\gamma f_2)}$, the decay rates of $J/\psi$ and $\Upsilon$ are simplified.
For $J/\psi\rightarrow\gamma f_2$ the value of $m_c$ makes \(T_2\sim 0\)
and \(A_2=0\) is taken. For $\Upsilon\rightarrow\gamma f_2$
because of small $x^2$ and $y^2$ then $T_1$ and $T_2$ are ignored. In QCD $J/\psi,\;\Upsilon\rightarrow light\;
hadrons$ are described as $J/\psi,\Upsilon\rightarrow 3g$ whose decay width is proportional to $\alpha_s^3 m_V$,
where $m_V$ is the mass of $J/\psi, \Upsilon$ respectively. The ratio is expressed as
\begin{eqnarray}
R=\frac{B(\Upsilon\rightarrow\gamma f_2)}{B(J/\psi\rightarrow\gamma f_2)}
=\frac{\Gamma(\Upsilon\rightarrow\gamma f_2)}{\Gamma(J/\psi\rightarrow\gamma f_2)}\nonumber
\end{eqnarray}
\begin{eqnarray}
\times\frac{\Gamma(J/\psi\rightarrow lh)}
{\Gamma(\Upsilon\rightarrow lh)}
\frac{B(\Upsilon\rightarrow lh)}{B(J/\psi\rightarrow lh)} \nonumber 
\end{eqnarray}
\begin{eqnarray}
=1.06\frac{\alpha_s(m_c)}{\alpha_s(m_b)}\frac{p^4_\Upsilon}{p^4_J}\nonumber 
\end{eqnarray}
\begin{eqnarray}
\frac{m_J m^6_c}{m_\Upsilon m^6_b}\frac{[m^2_c+{1\over4}(m^2_J-2m^2_f)]^2}
{[m^2_b+{1\over4}(m^2_\Upsilon-2m^2_f)]^2}
\frac{(1-{m^2_f\over m^2_\Upsilon})}{(1-{m^2_f\over m^2_\Upsilon})}\nonumber
\end{eqnarray}
\begin{eqnarray}
\frac{\{2m^2_f-m_\Upsilon(m_\Upsilon-2m_b)\}^2}{\{2m^2_f-m_J(m_J-2m_c)\}^2
+6{m^2_f\over m^2_J}\{m^2_f-{1\over2}(m_J-2m_c)^2\}^2},
\end{eqnarray}
where
\begin{eqnarray} 
p_\Upsilon={m^2_\Upsilon\over2m_f}(1-{m^2_f\over m^2_\Upsilon}),\nonumber \\
p_J={m^2_J\over2m_f}(1-{m^2_f\over m^2_J}).
\end{eqnarray}
Two factors affect the quark mass dependence of the ratio R.
It is the same as
$B(\Upsilon\rightarrow\gamma\eta'(\eta))$,
in the amplitude $(\Upsilon, J/\psi)\rightarrow\gamma gg, gg\rightarrow f_2$
$(\Upsilon, J/\psi)\rightarrow\gamma gg$ is strongly suppressed by heavy quark mass.
The second one is that d-wave dominates both the decay amplitudes of $(\Upsilon, J/\psi)\rightarrow\gamma f_2$.
The s-wave is only related to $A_2$ which is very small and
can be ignored in both cases. The enhancement factor,
$\frac{p^4_\Upsilon}{p^4_J}$, is the consequence of the d-wave dominance. This factor increases with
b-quark mass dramatically.
The combination of these two effects makes the dependence of
$\frac{B(\Upsilon\rightarrow\gamma f_2)}{B(J/\psi\rightarrow\gamma f_2)}$ on quark masses much weaker than the ratio
$\frac{B(\Upsilon\rightarrow\gamma \eta'(\eta)}{B(J/\psi\rightarrow\gamma\eta'(\eta))}$.
d-wave dominance in $\Upsilon,\; J/\psi\rightarrow\gamma f_2$ leads to larger $B(\Upsilon\rightarrow\gamma f_2)$.

\(m_c=1.29GeV\) and
\(m_b=(5.04\pm0.075\pm0.04)GeV\) are chosen to fit the experimental data of
$B(\Upsilon(1S)\rightarrow\gamma f_2(1270))$(3). The $m_c$ agrees with experimental value and $m_b$ is in agreement with
1S mass of $m_b$.







\end{document}